\documentclass[prb,twocolumn,showpacs,preprintnumbers,amsmath,amssymb]{revtex4}
\usepackage{graphicx}
\usepackage{dcolumn}
\usepackage{bm}

\begin{document}

\title{Towards New Half-Metallic Systems: Zinc-Blende Compounds of
  Transition Elements with N, P, As, Sb, S, Se, and Te}

\author{Iosif Galanakis}
\author{Phivos Mavropoulos}\email{Ph.Mavropoulos@fz-juelich.de}

\affiliation{Institut f\"ur Festk\"orperforschung, Forschungszentrum
  J\"ulich, D-52425 J\"ulich, Germany}

\date{\today}

\begin{abstract}
  We report systematic first-principles calculations for ordered
  zinc-blende compounds of the transition metal elements V, Cr, Mn
  with the $sp$ elements N, P, As, Sb, S, Se, Te, motivated by recent
  fabrication of zinc-blende CrAs, CrSb, and MnAs. They show
  ferromagnetic half-metallic behavior for a wide range of lattice
  constants. We discuss the origin and trends of half-metallicity,
  present the calculated equilibrium lattice constants, and examine
  the half-metallic behavior of their transition element terminated
  (001) surfaces.
\end{abstract}


\maketitle

\section{Introduction}

The swiftly developing field of spin electronics\cite{Prinz98} has
provided strong motivation towards the construction of novel magnetic
materials. Of particular interest are half-metals, {\it
  i.e.}~compounds for which only the one spin direction presents a gap
at the Fermi level $E_F$, while the other has a metallic character.
Such behavior is present in various perovskite
structures,\cite{Soulen98} in Heusler compounds,\cite{deGroot83} and
in dilute magnetic semiconductors.\cite{Matsukura98} Lately it has
been possible to grow binary CrAs in the zinc-blende (z-b) structure
epitaxially on GaAs; this compound was found to be half metallic, both
by experiment and by relevant calculations.\cite{Akinaga00} It has
also the great advantage of a high Curie temperature $T_C$, around
400~K.\cite{Mizuguchi02b} Similar is the $T_C$ of ferromagnetic z-b
CrSb.\cite{Zhao01} Moreover, the growth of nanoscale z-b MnAs dots on
GaAs substrates\cite{Ono02} has been achieved and CrAs/GaAs
multilayers have been fabricated.\cite{Mizuguchi02} Evidently, the
know-how on the construction of these materials, which perhaps can be
viewed as the limiting cases of dilute magnetic semiconductors, is
increasing, and they will possibly be at the center of interest of
spintronics applications soon, since they combine half-metallicity,
high Curie temperature, and coherent growth on semiconductor (SC)
substrates.

The experimental work on these systems seems to lie ahead of the
theoretical analysis. Pioneering calculations concerning the bulk or
the surfaces have been
reported,\cite{Akinaga00,Sanvito00,Shirai01,Continenza01,Zhao02,Galanakis02d,Liu02}
but there is still extensive work to be done. One matter to be
discussed is the origin of the half-metallicity; another is the
question whether half-metallicity is preserved at the surfaces; and a
third, in view of the future possible growth of similar compounds
based on different SC or transition metals, is the theoretical
prediction of the cases where half-metallicity can manifest itself, in
order to serve as a helpful guideline to where future experiments can
focus. It is the purpose of this paper to address these three points
by first-principles calculations.

The outline of the paper has as follows. In Section~\ref{Sec:calc} we
briefly describe our method of calculation. In Section~\ref{Sec:ES} we
address the question of the origin of the half-metallic behavior and
discuss the electronic structure of the z-b compounds of transiton
elements with $sp$ atoms; we examine only the ferromagnetic
configuration in view of the conclusions of Shirai\cite{Shirai01} on
VAs, CrAs, and MnAs. In Section~\ref{Sec:Trends} we present the
results of extensive systematic calculations for the z-b compounds of
the form XY, with X=V, Cr, Mn, and Y=N, P, As, Sb, S, Se, Te, for
various lattice constants in a ferromagnetic configuration, and we
examine in which cases half-metallicity is present; we
  also present the calculated equilibrium lattice constants to obtain
  a feeling for the lattice mismatch if one would grow these compounds
  on semiconductors. In Section~\ref{Sec:Surf} we focus on the
half-metallicity of transition metal terminated (001) surfaces.  We
conclude with a summary in Section~\ref{Sec:Summary}.

\section{Method of calculation\label{Sec:calc}}
To perform the calculations, we have used the Vosko, Wilk and Nusair
parameterization\cite{Vosko} for the local density approximation (LDA)
of the exchange-correlation potential to solve the Kohn-Sham equations
within the full-potential screened Korringa-Kohn-Rostoker (KKR) Green
function method,\cite{Papanikolaou02} where the correct shape of the
Wigner-Seitz cells is accounted for.\cite{Stefanou90} The scalar
relativistic approximation was used, which takes into account the
relativistic effects except that of spin-orbit
coupling.\cite{Bluegel87} To calculate the charge density, we
integrate along a contour on the complex energy plane, which extends
from below the $4s$ states of the $sp$ atom up to the Fermi level,
using 42 energy points. For the Brillouin zone (BZ) integration we
have used a $\mathbf{k}$-space grid of 30$\times$30$\times$30 in the
full BZ (752 $\mathbf{k}$-points in the irreducible wedge) for the
bulk calculations and a $\mathbf{k}_{\parallel}$-space grid
30$\times$30 (256 $\mathbf{k}_{\parallel}$-points in the irreducible
wedge) in the two-dimensional full BZ for the surface calculations.
We have used a cutoff of $\ell_{\mathrm{max}}=3$ for the wavefuctions
and Green functions. More details on the surface calculations are
described in Ref.~\onlinecite{Galanakis02d}.

The z-b structure can be easily seen to fit on a bcc lattice, by
occupying only half of the sites with the two kinds of atoms and
leaving the other half as voids. Viewed in this way, the consecutive
lattice sites of bcc in the cubic diagonal are occupied by Ga, As,
Void1, Void2 (in the case of {\it e.g.}~GaAs). This is useful for the
KKR method because all empty space of the quite open z-b structure is
taken into account.

We must also say a few words about state-counting with the KKR method.
The charge is found by adding the partial contributions of $s$, $p$,
$d$, and $f$ electrons (if a cutoff of $\ell_{\mathrm{max}}=3$ is
used). In this way one ignores the contribution to the wavefunction of
higher angular momenta, which, although very small, is nonzero. This
has a negligible effect for calculations in a metal, where the missing
(or extra) charge is regained by a very small readjustment of the
Fermi level $E_F$.  However, if there is a band gap at $E_F$, one
finds the charge to deviate slightly from the integer value it should
have by state-counting. As a consequence for half metals, altough the
densitiy of states (DOS) and the inspection of the bands show clearly
that $E_F$ lies within the gap for minority spin, and hence the
magnetic moment per unit cell must be integer, its calculated value
deviates slightly from the exact integer value.  In our work, the
presence of half metallicity has always been judged by inspection of
the density of states and of the energy bands, and not from the value
of the magnetic moment. The partial values of the charges and moments,
calculated by local orbital summation, are accurate to within 1\%.

\section{Electronic structure\label{Sec:ES}}
In this section we shall examine the electronic structure of the
compounds, and discuss the origin of the half metallicity and the
value of the total moment. Although for arbitrary $\mathbf{k}$ the
wavefunctions do not belong exclusively to a single irreducible
representation such as $t_{2g}$ or $e_g$, it is convenient to retain
this terminology for the bands formed by originally $t_{2g}$ or $e_g$
orbitals since they are energetically rather separated. Let
us focus on a $3d$ atom.  In the z-b structure, the tetrahedral
environment allows its $t_{2g}$ states ($d_{xy}$, $d_{yz}$, and
$d_{xz}$) to hybridize with the $p$ states of the four first neighbors
(the $sp$ atoms). Note that in the tetrahedral geometry, the $p$
orbitals of the four neighboring sites transform as partners of the
$t_{2g}$ irreducible representation when analyzed around the central
site.\footnote{Again, this is strict only for $\mathbf{k}=0$} This
creates a large bonding-antibonding splitting, with the low-lying
bonding states being more of $p$ character around the $sp$ neighbors
(we call these bands ``$p$-bands'' henceforth), and the antibonding
being rather of $d$ character around the $3d$ atom (but see
  also the discussion below concerning Refs.~\onlinecite{Sanvito00}
  and \onlinecite{Continenza01}). The gap formed in between
is partly filled by the $e_g$ states of the $3d$ atom ($d_{z^2}$ and
$d_{x^2-y^2}$); the position of the gap will be different for majority
and minority electrons, due to the exchange splitting.

This symmetry-induced $p$-$d$ hybridization and bonding-antibonding
splitting has been found in the past in the case of transiton metal
doped zinc-blende semiconductors;\cite{Zunger,Wei87} the splitting has
also been refered to as ``$p$-$d$ repulsion'', concerning the
interaction of the impurity $t_{2g}$ localized states with the
semiconductor anion $p$ band. In the limit of full substitution of the
semiconductor cation by the transition metal, one arrives to the cases
studied here; in particular MnTe has been studied in
Ref.~\onlinecite{Wei87}, and has been found half-metallic if the
ferromagnetic phase is assumed.

\begin{figure}[t]
\begin{center}
\includegraphics[scale=0.4]{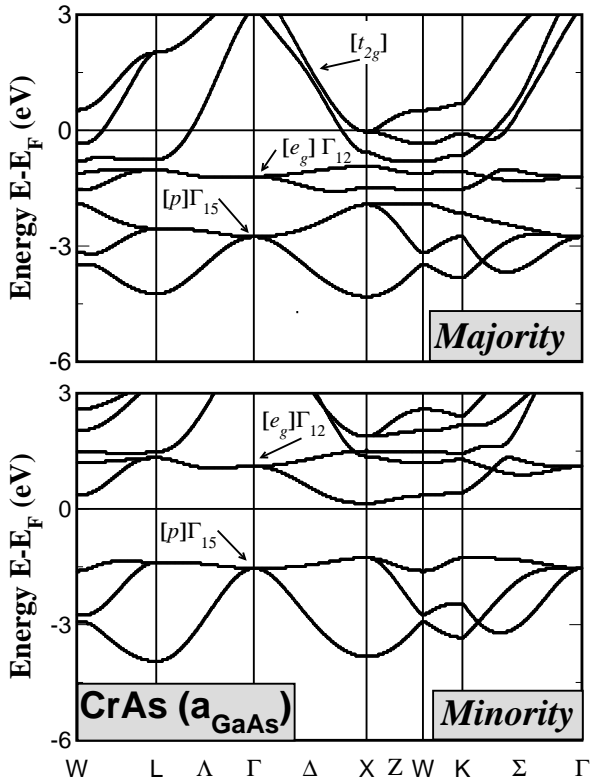}
\end{center}
\caption{Energy bands of CrAs calculated at the GaAs lattice
  constant. With $p$, $e_g$, and $t_{2g}$ we denote the whole bands
  originating from orbitals of the corresponding character, although
  this character is well defined only for $\mathbf{k}=0$. The
  bonding-antibonding splitting between the $p$ and $t_{2g}$ bands is
  evident; the band gap is between the $p$ and the flat $e_g$ bands.
  The system is half-metallic. }
\label{fig1}
\end{figure}

The situation can be elucidated by examination of the energy bands of
CrAs, presented in Fig.~\ref{fig1}.  The $s$ states of As are very low
in energy and omitted in the figure. Around $-3\ \mathrm{eV}$, the
$p$-bands can be seen. The next bands are the ones formed by the Cr
$e_g$ orbitals, around $-1.5 \ \mathrm{eV}$ for majority and $+1\ 
\mathrm{eV}$ for minority. They are quite flat, reflecting the fact
that their hybridization with the states of As neighbors is weak (or
even zero, at $\mathbf{k}=0$, due to symmetry). These bands can accommodate
two electrons per spin. Above them, the substantially wider
antibonding $p$-$t_{2g}$ hybrids appear, starting from $-1
\mathrm{eV}$ for majority and from $+1.5\ \mathrm{eV}$ for minority.
Their large bandwidth can be attributed to the strong hybridization
and to their high energy position. The bonding-antibonding splitting
is also contributing to the fact that they be pushed up beyond the
flat $e_g$ states. As a result, the three families of bands do not
intermix, but are rather energetically separate.
In the majority-spin direction the bonding-antibonding splitting is
smaller, because the transition element $3d$ states are originally
lower in energy, closer to the $p$ states of the $sp$ atom. In the
minority spin direction the states are higher due to the exchange
splitting and the gap is around $E_F$. Band counting gives an integer
total magnetic moment of 3 $\mu_B$.

To the clear bonding-antibonding gap also the z-b geometry
contributes, since it has been previously
reported\cite{Sanvito00,Continenza01,Zhao02} that in the hexagonal
NiAs geometry MnP, MnAs and MnSb show no gap.

\begin{figure}[t]
\includegraphics[scale=0.62,angle=270]{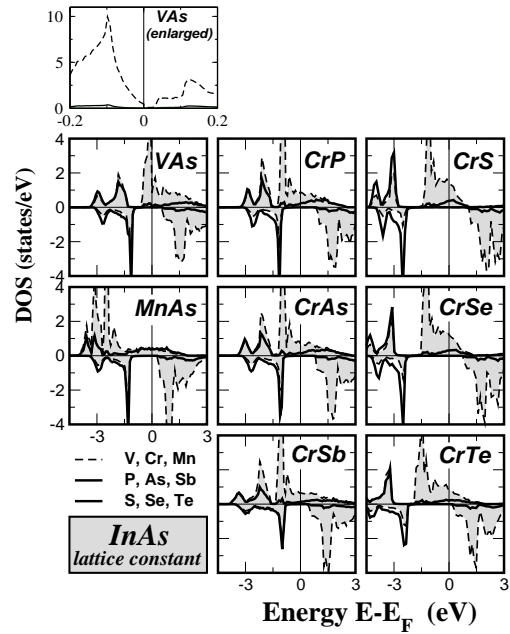}
 \caption{Atom-resolved DOS for several materials in the z-b structure using the
   InAs lattice constant. Negative numbers on the DOS axis represent
   the minority spin. The dashed lines containing the shadowed area
   correspond to the DOS within the Wigner-Seitz cell of the
   transition element, while the full lines to the one of the $sp$
   atom. The small DOS of the vacant sites is omitted. In all cases,
   $E_F$ lies within the minority-spin gap, {\it i.e.}~the systems are
   half-metallic. In the upper left inset, the majority DOS for VAs is
   shown in more detail around $E_F$, to demonstrate the low DOS at
   $E_F$. Spin moments for these compounds are given in
   Table~\protect{\ref{Table:3}}.}
\label{fig2}
\end{figure}

Here we must refer to two recent publications on zinc-blende
MnAs\cite{Sanvito00}, and MnP, MnAs, and MnSb,\cite{Continenza01}
which have identified the $p$ and $t_{2g}$ bands the other way around;
{\it i.e.}~they state that the bonding bands are more of $d$-character
around Mn and the antibonding more of $p$ character around the $sp$
atom. This is correct for the majority bands in those compounds where
the majority transition-element $d$ states are originally lower than
the anion $p$ states, {\it e.g.}~in the case of the MnAs or CrAs
bonding majority bands. For minority-spin the situation is the other
way around, so that the bonding bands are more $p$-like centered
around the $sp$ cation, while the antibonding bands are more $d$-like
centered around the transition element. This can be seen by the
atom-resolved DOS in Fig.~\ref{fig2}. For the group-VI compounds, the
$p$ states are initially even lower, so that the $p$ character of the
bonding states increases for both spins.

Next we consider different compounds, to see the trend of increasing
transition metal atomic number: VAs, CrAs, MnAs. The corresponding DOS
can be seen in Fig.~\ref{fig2}, calculated for the InAs lattice
parameter in all cases to allow comparison. For VAs, $E_F$ lies
between the majority $e_g$ and $t_{2g}$. Since they are separated,
this results in zero DOS at $E_F$, as in the limiting case of a
semiconductor with almost zero band gap (shown in more detail in the
corresponding inset of Fig.~\ref{fig2}). The moment is $2\ \mu_B$. As
the number of valence $3d$ electrons increases for CrAs, the $t_{2g}$
bands are pulled down enough to accommodate one electron ($M=3\ 
\mu_B$), and for MnAs even more, so as to accommodate two electrons,
whence $M=4\ \mu_B$. At the same time, for the minority spin the band
gap does not change in magnitude and continues to embrace $E_F$. This
can be understood in terms of the exchange splitting $\Delta E_x$
being proportional to the magnetic moment: $\Delta E_x \simeq I\,M$.
When the moment increases by 1 $\mu_B$, $\Delta E_x$ must increase by
one time the exchange integral $I$, about $0.8\ 
\mathrm{eV}$.\cite{Janak77} Thus the minority bands remain almost at
their position.

This behavior is quite stable, unless the energy gained by
increasing the exchange splitting in order to achieve one more
$\mu_B$ is overshot by the Coulomb energy cost; remember that the
majority $t_{2g}$ bands are wide. Then the system compromises by
placing $E_F$ higher, within the minority $e_g$ conduction band,
and half metallicity is lost. This can occur for small lattice
constants or at the surface, and will be presented in the next
sections.

Generally, the bonding $p$ bands can accommodate 6 electrons (3 per
spin), and since they lie low in energy they must be filled up; in
addition, there is yet another band lower in energy, consisting of the
$s$ states of the $sp$ atom, which has space for two electrons (one
per spin). Once these 8 bands are filled by 8 of the valence
electrons, the remaining electrons will fill part of majority $d$
states, first the lower-lying $e_g$ and then the $t_{2g}$, creating
the magnetic moment. This electron counting gives then an integer
total moment of
\begin{equation}
M=(Z_{\mathrm{tot}}-8)\ \mu_B,
\label{eq:SP}
\end{equation}
where $Z_{\mathrm{tot}}$ is the total number of valence electrons in
the unit cell.\footnote{The cumbersome electron counting can be
  avoided by just observing that $M=N_{\mathrm{maj}}-N_{\mathrm{min}}=
  N_{\mathrm{maj}}+N_{\mathrm{min}}-2\times N_{\mathrm{min}} =
  Z_{\mathrm{tot}}-2\times N_{\mathrm{min}}$, where $N_{\mathrm{maj}}$
  and $N_{\mathrm{min}}$ are respectively the numbers of majority and
  minority valence electrons.} This ``rule of 8'' is the analogue of
the Slater-Pauling behavior for these systems; similar behavior is
found in the case of the half-metallic Heusler alloys, as a ``rule of
18'' or a ``rule of 24''.\cite{Galanakis02a,Galanakis02b} Of
course these are valid only when half-metallicity is present.

According to the ``rule of 8'' of Eq.~(\ref{eq:SP}), the (integer)
moment should increase by $1\ \mu_B$ if one changes only the $sp$ atom
from group V to group VI, since then one more valence electron is
available (in this respect there is a similarity to the dilute
magnetic semiconductors\cite{Sato02}). In order to present this, we
calculated the electronic structure of the $3d$--group VI compounds.
In Fig.~\ref{fig2} the DOS of CrS, CrSe, and CrTe are shown.  We chose
to use the InAs lattice constant in this case, so as to compare with
the group V compounds at a single lattice constant.  Note that the
InAs lattice (6.058\AA ) is very close to the CdSe one (6.052\AA ).
We found a ferromagnetic half-metallic solution for these cases, with
a magnetic moment of 4 $\mu_B$ as expected. The physical mechanism
involved is the same as in the $3d$--group V case, and the main
difference lies in that $p$ states of the group VI elements lie lower
in energy, so that the band gap is larger and the hybridization with
the neighbor $d$ states smaller.

We have also calculated the $3d$--group VI compounds VS, VSe, VTe,
MnS, MnSe, and MnTe. They also show half-metallic behavior for a range
of lattice constants, with their magnetic moment given by the ``rule
of 8''; they will be discussed in the next section. However, we must
note that we have only searched for the ferromagnetic solution. A
systematic investigation of the energetics of ferromagnetic {\it
  vs.}~antiferromagnetic solutions lies beyond the scope of this
paper.  Especially in the case of Mn--group VI compounds, an
antiferromagnetic solution could be possibly more stable, since these
compounds are isoelectronic with FeAs, which has been found to have an
antiferromagnetic ground state at the equilibrium lattice constant in
calculations by Shirai;\cite{Shirai01} especially MnTe has been
calculated to be more stable in the antifferomagnetic phase, in
Ref.~\onlinecite{Wei87}. For the compounds VAs, CrAs, and MnAs, the
analysis of Shirai points towards stable ferromagnetic solutions.

\begin{table}
\caption{\label{Table:3} Spin magnetic moment in $\mu_B$ for the
  InAs experimental lattice constant. The atomic-resolved moment
  values refer to the moment included in a Wigner-Seitz cell around
  each atom. Void1 and Void2 refer to the vacant sites used to
  describe the zinc-blende structure as described in
  Section~\protect{\ref{Sec:calc}}. The total moment does not appear
  exactly integer as explained in Section~\protect{\ref{Sec:calc}}.}
\begin{ruledtabular}
\begin{tabular}{rrrrrr}
 Compound &  $3d$at.   & $sp$-at.  &  Void1  & Void2 & Total\\ 
\hline
CrN  & 3.887 & -1.071 &  0.027 & 0.127 & 2.960 \\
CrP  & 3.318 & -0.448 & -0.019 & 0.085 & 2.935 \\
CrAs & 3.269 & -0.382 & -0.029 & 0.080 & 2.937 \\
CrSb & 3.148 & -0.249 & -0.036 & 0.084 & 2.946 \\
VAs  & 2.054 & -0.196 &  0.006 & 0.076 & 1.939 \\
MnAs & 4.070 & -0.250 &  0.008 & 0.106 & 3.935 \\
CrS  & 3.858 & -0.117 &  0.050 & 0.154 & 3.945 \\
CrSe & 3.830 & -0.102 &  0.050 & 0.162 & 3.939 \\
CrTe & 3.752 & -0.057 &  0.053 & 0.182 & 3.930 \\
VSe  & 2.750 & -0.054 &  0.068 & 0.173 & 2.937 \\
MnSe & 4.579 &  0.149 &  0.061 & 0.142 & 4.931
\end{tabular}
\end{ruledtabular}
\end{table}

In Table~\ref{Table:3} the magnetic moments are presented for several
$3d$--group V and $3d$--group VI materials for the InAs lattice
constants.  All these compounds are half-metallic, even MnSe, where
the $t_{2g}$ majority spin band is filled completely to achieve the 5
$\mu_B$.  The total moment, calculated by summation over KKR local
orbitals instead of state counting, does not appear exactly integer as
it should, as explained in Section~\protect{\ref{Sec:calc}}.

From Table~\ref{Table:3} we see that the $sp$ atom has an induced
local magnetic moment $M_{\mathrm{loc}}$ opposite to the one of the
$d$ atom. This is in agreement with previous results on $3d$--As
compounds.\cite{Sanvito00,Shirai01,Continenza01} The trend is that for
lighter $sp$ elements the induced moment absolute value increases
($M_{\mathrm{loc}}$ becomes more negative); naturally, the local
moment at the $3d$ atom increases as well, since the total magnetic
moment must remain a constant integer. For the group VI elements
(higher ionicity) these effects are smaller.

This behavior can be understood in the following way. The minority
bonding $p$ band is more located within the Wigner-Seitz cell of the
$sp$ atom, whereas the majority $p$ band is more hybridized with the
low $d$ orbitals and thus more shared with the $3d$ atoms; this
results in the opposite magnetic moments of the $sp$ atoms. As we
change to lighter $sp$ elements, the minority $p$ wavefunctions become
even more localized, while the majority ones are less affected due to
their strong hybridisation with the low $d$ orbitals (this is evident
in the atomic resolution of the DOS of CrSb, CrAs, and CrP in
Fig.~\ref{fig2}). Thus the absolute value of the local moment on the
$sp$ atoms --- antiparallel to the one of the $3d$ atoms --- increases
for the lighter elements.  In the case of $3d$--Group VI compounds,
these effects are smaller, because the $p$ states are deeper in
energy, thus less hybridized with the transition element majority $d$
states and more alike for the two spin directions.

Another effect of going to lighter $sp$ elements is the increase of
the band gap, as can be seen in Fig.~\ref{fig2} for
CrSb$\rightarrow$CrAs$\rightarrow$CrP and
CrTe$\rightarrow$CrSe$\rightarrow$CrS. This is consistent with the
picture that the local magnetic moment of the $3d$ atom increases,
causing a larger exchange splitting $\Delta E_x$ and pushing up the
minority $e_g$ bands.

\section{Variation of lattice constant\label{Sec:Trends}}

The compounds consisting of $3d$ and $sp$ elements that have been
fabricated\cite{Akinaga00,Zhao01,Ono02} are metastable in the z-b
structure. For instance, MnAs is grown in the orthorombic or hexagonal
structure,\cite{Kaganer00} in which it does not present half-metallic
behavior,\cite{Sanvito00,Continenza01} although it is still
ferromagnetic. The metastable z-b structure can be achieved only when
the materials are grown on top of a z-b semiconductor, such as
GaAs.\cite{Akinaga00,Zhao01,Ono02} It can be expected that in the
first few monolayers the materials will adopt the lattice constant of
the underlying buffer; thus, the question arises whether
half-metallicity is present for the lattice parameter of the
semiconductor. To peruse this, we have performed calculations of all
combinations of the type XY with X=V, Cr, Mn, and Y=N, P, As, Sb, for
the lattice constants of the relevant z-b III-V semiconductors GaN,
InN, GaP, GaAs, InP, InAs, GaSb, and InSb (given in increasing lattice
parameter order). 

Moreover, we have calculated the equilibrium lattice constant that
these compounds would have. In this manner one can obtain a feeling of
what systems would perhaps grow on a certain SC substrate, as opposed
to unrealistic combinations due to lattice mismatch and stress. Here
we must note that within the LDA, the equilibrium lattice constant is
known to be underestimated by a few percent\cite{Asato99} (the
so-called overbinding); thus the ``real'' lattice constant of such
materials would be as much as 3\% larger.

Our results are summarized in Table~\ref{Table:1}. A ``$+$'' sign
means that the compound is half-metallic for the given lattice
constant, while a ``$-$'' sign means loss of half-metallicity. A $\pm$
means that though the system is not half metallic, the spin
polarization at $E_F$ is very close to 100\%. Where half-metallicity
is preserved, the total magnetic moment is given by Eq.~(\ref{eq:SP}).
Note that even in the cases where half-metallicity is lost, a $p$-$d$
hybridization gap is present (but not at $E_F$). Upon compression, the
bandwidth of the valence $p$-bands and the $e_g$ bands increases and
the gap between them decreases somewhat --- the increase of the
bonding-antibonding splitting is not relevant, since it concerns the
relative position of the $p$ and $t_{2g}$ bands. Most importantly,
though, $E_F$ is shifted towards higher energies by the majority
states, so that it can step into the minority $e_g$ band and destroy
half-metallicity.  In Table~\ref{Table:1} we see that all materials
examined are half-metallic for the large lattice constant of InSb, but
lose the half-metallicity for the small one of GaN.

\begin{table}
\caption{\label{Table:1} Calculated properties of different
  materials adopting the zinc-blende structure using the experimental
  lattice constants a of several z-b III-V semiconductors. A ``$+$''
  sign means that a system is half-metallic, a ``$-$'' that the Fermi
  level is above the gap and a ``$\pm$'' that the Fermi level is only
  slightly within the conduction band and the spin polarization at
  $E_F$ is almost 100\%. For each compound, the calculated equilibrium
  lattice parameter is also given in parentheses. The experimental
  semiconductor lattice parameters were taken from
  Refs.~\protect{\onlinecite{landolt1,landolt2}} except for the case
  of GaN and InN in the zinc-blende structure which were taken from
  Refs.~\protect{\onlinecite{landolt1,landolt3}}.}
\begin{ruledtabular}
\begin{tabular}{lcccccccc}
 a(\AA )& GaN & InN & GaP  & GaAs & InP & InAs & GaSb & InSb \\
 Compound & 4.51 & 4.98 & 5.45 & 5.65 & 5.87 & 6.06 & 6.10 & 6.48 \\ 
\hline
 VN   (4.21)& $-$ & $-$ & $+$ & $+$ & $+$ & $+$ & $+$ & $+$ \\
 VP   (5.27)& $-$ & $-$ & $-$ & $+$ & $+$ & $+$ & $+$ & $+$ \\
 VAs  (5.54)& $-$ & $-$ & $-$ & $+$ & $+$ & $+$ & $+$ & $+$ \\
 VSb  (5.98)& $-$ & $-$ & $-$ & $-$ & $+$ & $+$ & $+$ & $+$ \\
 CrN  (4.08)& $-$ & $+$ & $+$ & $+$ & $+$ & $+$ & $+$ & $+$ \\
 CrP  (5.19)& $-$ & $-$ & $-$ & $+$ & $+$ & $+$ & $+$ & $+$ \\
 CrAs (5.52)& $-$ & $-$ & $-$ & $+$ & $+$ & $+$ & $+$ & $+$ \\
 CrSb (5.92)& $-$ & $-$ & $-$ & $-$ & $+$ & $+$ & $+$ & $+$ \\
 MnN  (4.06)& $-$ & $+$ & $+$ & $+$ & $+$ & $+$ & $+$ & $+$ \\
 MnP  (5.00)& $-$ & $-$ & $-$ & $-$ & $+$ & $+$ & $+$ & $+$ \\
 MnAs (5.36)& $-$ & $-$ & $-$ & $-$ & $+$ & $+$ & $+$ & $+$ \\
 MnSb (5.88)& $-$ & $-$ & $-$ & $-$ & $-$ & $-$ & $\pm$ & $+$
\end{tabular}
\end{ruledtabular}
\end{table}

The general trend observed in Table~\ref{Table:1}, apart from the
effect of compression, is that the tendency towards half-metallicity
increases for lighter $sp$ elements. This is consistent with the
picture that for lighter group V elements the $p$ state is
energetically lower and also more localized. Both these factors result
in reducing the shifting of $E_F$ by compression; but more
importantly, the gap increases for lighter $sp$ elements as analyzed
at the end of Sec.~\ref{Sec:ES}. Thus half-metallicity holds out
longer.

\begin{table}
\caption{\label{Table:2}Similar to Table~\protect{\ref{Table:1}} 
  for transition metal compounds with the group VI elemetns.  For each
  compound, the calculated equilibrium lattice parameter is also given
  in parentheses. The experimental semiconductor lattice parameters
  were taken from Ref.~\protect{\onlinecite{landolt1}}.}
\begin{ruledtabular}
\begin{tabular}{lcccccc}
 a(\AA )& ZnS & ZnSe & CdS  & CdSe & ZnTe & CdTe \\
 Compound & 5.41 & 5.67 & 5.82 & 6.05 & 6.10 & 6.49 \\ 
\hline
 VS   (5.24)& $-$ & $-$  & $+$  & $+$ & $+$  & $+$ \\
 VSe  (5.56)& $-$ & $-$  & $-$  & $+$ & $+$  & $+$ \\
 VTe  (6.06)& $-$ & $-$  & $-$  & $-$ & $\pm$& $+$ \\
 CrS  (5.04)& $-$ & $+$  & $+$  & $+$ & $+$  & $+$ \\
 CrSe (5.61)& $-$ & $\pm$& $+$  & $+$ & $+$  & $+$ \\
 CrTe (6.07)& $-$ & $-$  & $-$  & $+$ & $+$  & $+$ \\
 MnS  (4.90)& $-$ & $+$  & $+$  & $+$ & $+$  & $+$ \\
 MnSe (5.65)& $-$ & $-$  & $\pm$& $+$ & $+$  & $+$ \\
 MnTe (6.10)& $-$ & $-$  & $-$  & $-$ & $-$  & $+$ 
\end{tabular}
\end{ruledtabular}
\end{table}

Similar trends are observed in the case of $3d$--group VI compounds.
The systematics are presented in Table~\ref{Table:2}. The investigated
systems are XY with X=V, Cr, and Mn, and Y=S, Se, and Te, for the
experimental lattice constants of the II-VI semiconductors ZnS, ZnSe,
CdS, CdSe, ZnTe, and CdTe. Again we see the loss of half-metallicity
by compression, and a stronger tendency towards half-metallicity for
lighter group VI elements; both effects can be attributed to the same
reasons as in the $3d$--group V case.

However, the calculated equilibrium lattice constants (given in
parentheses in Tables~\ref{Table:1} and \ref{Table:2}) lessen the
possible selections. Indeed, the nitrides and sulphides tend to be so
much condensed that they would probably not grow epitaxially at a
lattice constant favouring half-metallicity; the nitrides are even
nonmagnetic at their small equilibrium lattice constant. In fact, only
VSb, CrSb and CrTe are half-metallic at their equilibrium lattice
constant. On the other hand, if one would allow for a 5\% expansion
due to the LDA overbinding or tollerable lattice mismatch, one has the
possible combinations VAs/GaAs, VSb/InAs, CrAs/GaAs, CrSb/InAs,
VTe/CdTe, CrSe/CdS, CrTe/ZnTe.

Other authors have reported results on the equilibrium lattice
constant of MnAs, CrAs,\cite{Shirai01} VAs,\cite{Shirai01} and
CrSb.\cite{Liu02} For MnAs, the most frequently studied of these, the
reported LDA results are: 5.85~\AA (Ref.~\onlinecite{Shirai01}),
5.6-5.7~\AA (Ref.~\onlinecite{Sanvito00}), 5.45~\AA
(Ref.~\onlinecite{Zhao02}), and 5.34~\AA
(Ref.~\onlinecite{Continenza01}). Our calculations give 5.36~\AA .  We
do not know the reason for the discrepancies among the reported
values. Changing the $\ell_\mathrm{max}$ cutoff or the number of
$\mathbf{k}$-points brought only very small changes to our result,
while using non-relativistic treatment for the valence electrons
resulted in an increase to 5.41~\AA . Results using the generalised
gradient approximation (GGA) as an improvement to the LDA have been
reported in Refs.~\onlinecite{Continenza01}, \onlinecite{Zhao02}, and
\onlinecite{Liu02}, giving slightly larger lattice constants compared
with LDA.

\section{(001) Surfaces \label{Sec:Surf}}

In the last part of this paper we examine the possibility of
half-metallicity at the surfaces of these compounds.  One of the
authors (IG) has recently studied the (001) surface of
CrAs.\cite{Galanakis02d} The conclusion of that study was that the As
terminated surface loses the half-metallicity due to As dangling bonds
that form a surface band within the minority spin gap, while the Cr
terminated surface retains the half-metallic character and the local
magnetic moment increases.  This behavior was found for both the GaAs
and InAs lattice constants.  Here we have extended our calculations to
include the VAs and MnAs (001) surfaces, for the InAs lattice
constant. The atoms were kept in their ideal bulk-like positions, {\it
  i.e.}, we have calculated the unrelaxed and unreconstructed surface;
we must note that large relaxations or reconstruction could alter the
half-metallic character of the surface. Since for the As terminated
surfaces the surface states originating from dangling bonds cannot be
avoided, we will discuss the termination with the $3d$ atom. Although
at the surface the $t_{2g}$ and $e_g$ representations are no longer
relevant (the former splits in two, one containing $d_{xz}+d_{yz}$ and
one $d_{xy}$, and the latter also in two, one containing $d_{x^2-y^2}$
and one $d_{z^2}$), we ask for understanding that we continue to use
the same terminology in order to make the connection to the bulk
states.

\begin{figure}[t]
\includegraphics[scale=0.32,angle=270]{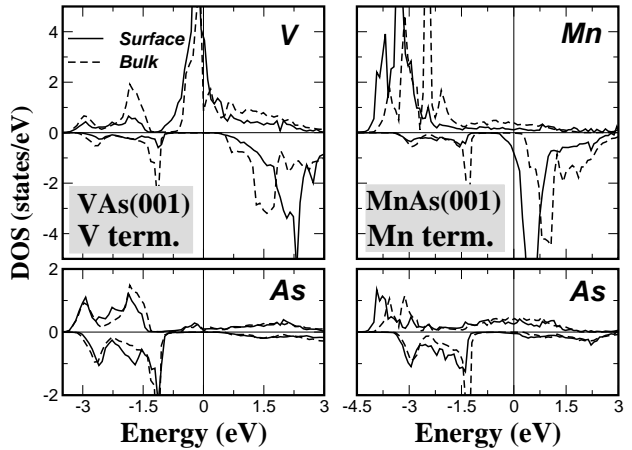}
 \caption{\label{fig3}Left pannel: Spin- and atom-resolved DOS of  the
   V-terminated (001) surface for the V atom at the surface layer and
   the As atom at the subsurface layer for the InAs lattice constant
   (left pannel). The surface DOS (full lines) and the bulk DOS
   (dashed lines) are shown. The As $s$ states are located at around
   10 eV below the Fermi level and are not shown in the figures. The
   energy zero refers to $E_F$; negative numbers on the DOS axis
   represent the minority spin.  Right pannel: Same for the
   Mn-terminated MnAs (001) surface.}
\end{figure}

The DOS for the VAs and MnAs (001) surfaces are shown in
Fig.~\ref{fig3}, for both the atom at the surface layer (V or Mn) and
at the subsurface layer (As).  For comparison, the corresponding DOS
in the bulk are also presented.  Evidently the V terminated VAs (001)
surface remains half-metallic.  However, there is an important
difference in the DOS as compared to the bulk case, especially at the
surface (V) layer. For the majority spin, the peak at around $-2\ 
\mathrm{eV}$ is strongly reduced, and the same happens for the
minority spin at around $-1\ \mathrm{eV}$.  This can be explained by
the fact that the surface V atom has a reduced coordination number of
only two As neighbors. Then two less As $p$ orbitals penetrate the V
cell, reducing the local weight of the bonding states. At the same
time, the majority peak just under the Fermi level is enhanced,
gaining the weight which has been lost by the bonding states from both
spin directions. Note in the DOS of the V surface atom that there is
no more the clear minimum at $E_F$ for the majority states, which was
distinguishing $e_g$ from $t_{2g}$ in the bulk, as the reduced
bonding-antibonding splitting allows the latter states to come lower
in energy.

\begin{figure}[t]
\includegraphics[width=8cm]{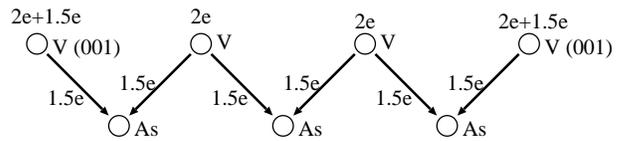}
\caption{Schematic interpretation of the origin of the excess moment
  of 1.5 $\mu_B$ at the VAs surface. The arrows represent the donation
  of V electrons to the valence band generated mainly by the low-lying
  As $p$ orbitals. At the V-terminated (001) surface, half the
  neighbors of the V atom are missing, so an excess of 1.5 electrons
  remains. They occupy the majority-spin states, giving an extra 1.5
  $\mu_B$. The situation is similar in all $3d$-group V compounds; for
  the $3d$-group VI compounds the $p$ band welcomes in total 2 instead
  of 3 electrons from each $3d$ atom, thus the excess moment is 1
  instead of 1.5 $\mu_B$.}
\label{figSchema}
\end{figure}

In terms of electron counting, the V atoms in the bulk give away 3 of
the 5 valence electrons to the $p$ band and retain the remaining two
to build up the magnetic moment. On the other hand, due to the reduced
coordination at the surface, only 1.5 electron per V atom is given
away, thus 3.5 remain to fill up majority states and build up the
magnetic moment; due to the gap, the minority states cannot take over any
of these extra electrons, unless $E_F$ moves higher into the
$e_g$-like band (this is the case for Mn to which we shall turn
shortly). The situation is given schematically in
Fig.~\ref{figSchema}.

The gain in magnetic moment at the surface is accompanied by an
increase of the exchange splitting, pushing the antibonding and
$e_g$-like minority peaks higher in energy (from 1.5 eV to 2 eV);
otherwise the gap could have decreased due to reduced hybridisation.

The increased magnetic moment must reach $3.5\ \mu_B$
in the vicinity of the surface; we find that $3.07\ \mu_B$ are located
on the surface layer, about $0.35\ \mu_B$ in the vacuum, and the rest
adds up to the moment of the subsurface layer. We see that the integer
value of the magnetic moment, which is mandatory in bulk
half-metallicity, ceases to be a prerequisite in the vicinity of the
surface.

The situation described here is completely analogous to the one of the
CrAs (001) surface.\cite{Galanakis02d} There, the magnetic moment of
the surface layer was reported to reach 4 $\mu_B$; if the magnetic
moment of the vacuum and the subsurface layer is added, one finds a
total moment of 4.5 $\mu_B$ in the vicinity of the surface, {\it i.e.}
1.5 $\mu_B$ more than in the bulk.

However, things are different for MnAs (001) (with Mn termination).
The corresponding DOS of the surface Mn atom and of the subsurface As
atom is shown in Fig.~\ref{fig3} (right panel). Manifestly,
half-metallicity has been lost. The difference to the VAs and CrAs
cases lies in the high magnetic moment of $4\ \mu_B$ that MnAs already
shows in the bulk. If its moment were to increase by the same
mechanism, the surface (plus vacuum and subsurface layers) would have
to accommodate 5.5 majority spin electrons. This is energetically
unfavorable, since the extra half electron cannot be accommodated by
the $t_{2g}$ band, so that higher-lying bands would have to be
populated. Instead, as the majority states are lowered, the slow gain
in magnetic moment is not enough to increase the exchange splitting
substantially, and the minority $d$ states also reach $E_F$.  As a net
result, majority and minority DOS peaks are substantially lower in
energy, half-metallicity is lost, and a total magnetic moment of
$4.59\ \mu_B$ is present within the first surface layer.

\begin{figure}[t!]
\includegraphics[scale=0.32,angle=270]{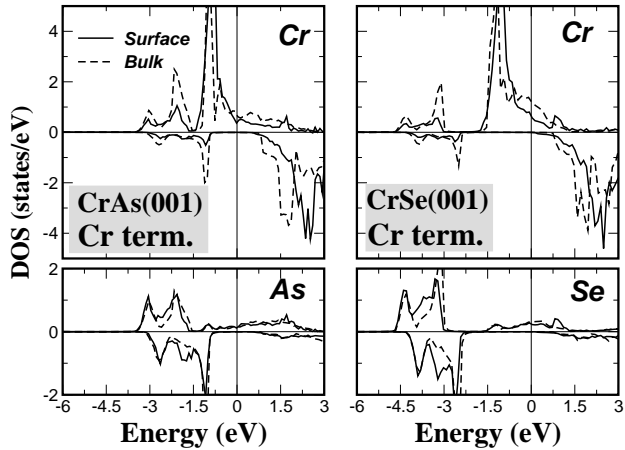}
 \caption{\label{fig4}Spin- and atom-resolved DOS of  the
   Cr-terminated CrAs and CrSe (001) surfaces for the Cr atom at the
   surface layer and the As and Se atoms at the subsurface layer for
   the InAs lattice constant. The As and Se $s$ states are located at
   around 10 eV below the Fermi level and are not shown in the
   figures. The surface DOS are compared to the bulk calculations
   (dashed lines). The energy zero refers to $E_F$; negative numbers
   on the DOS axis represent the minority spin.}
\end{figure}
We have also made calculations for the X terminated XSe(001) surfaces,
where X=V, Cr, and Mn, using the InAs lattice constant.  The V and Cr
ones are half-metallic while the Mn surface is not half-metallic.  The
arguments in the above paragraphs still hold, but now each V, Mn or Cr
atom loses Se (group VI) neighbors instead of As (group V) and thus
the electron counting changes.  In the bulk each X atom gives away
only 2 electrons to the $p$ band and not 3 as in the XAs compounds.
Losing two neighbors means that the surface X atom gives away now only
1 electron, so we should have now 4 electrons for V or 5 electrons for
Cr to build up the spin moment, {\it i.e.}~we should have 1 $\mu_B$
more than in the bulk. Indeed we find the total spin moment to
increase to 4 $\mu_B$ for VSe(001) and 5 $\mu_B$ for CrSe(001). The
MnSe compound cannot be half-metallic as this would mean a total spin
moment of 6 $\mu_B$ at the surface which is energetically unfavourable
for reasons similar to the MnAs surface.

A comparison of the CrAs(001) with the CrSe(001) surface DOS, both Cr
terminated and at the InAs lattice constant, is shown in
Fig.~\ref{fig4}, where the bulk DOS is also presented. We see that the
differences in the bulk of the two materials (larger gap, more weight
in the majority states under $E_F$ to account for one more electron)
remain their major differences also in the surfaces.

\section{Summary and Conclusion\label{Sec:Summary}}

We have studied zinc-blende compounds of the transition metal elements
V, Cr, Mn with the group V $sp$ elements N, P, As, Sb and the group VI
S, Se, Te, in the ferromagnetic configuration. They all show a
tendency towards half-metallic behavior, {\it i.e.}~100\% spin
polarization at $E_F$.  This can be traced back to the
bonding-antibonding splitting due to hybridization between the
transition element $d$ ($t_{2g}$) states and the $sp$ element $p$
states, conspiring with the large exchange splitting which pushes up
the minority $d$ states.  The total moment per unit cell, if the
system is half-metallic, is integer and given by the ``rule of 8''.
Also, the $sp$ atoms are found to have an antiparallel local moment
with respect to the $3d$ atoms; the absolute value of the local
moments increases for lighter or less ionized $sp$ elements. This is
traced back to the degree of localization of the $p$ wavefunctions
around the $sp$ atoms.

We have discussed the trends with varying lattice constant, in view of
the possibility to grow these materials epitaxially on various
semiconductors, and calculated the LDA equilibrium lattice constants
to obtain a feeling of the lattice mismatch with the possible SC
substrates.  Compression eventually kills half-metallicity, since
$E_F$ finally wanders above the minority gap. Although for compounds
involving lighter $sp$ elements half-metallicity is seems more robust,
there equilibrium lattice constants are too small.  Thus, the best
candidates for half-metallicity from this aspect are VAs, VSb, CrAs,
CrSb, VTe, CrAs, and CrTe.

Finally, we have examined the behavior of the transition element
terminated (001) surfaces. In most cases half metallicity is
maintained, and the magnetic moment increases because of the missing
neighbours where charge would be transfered. Exceptions are the cases
where the surface magnetic moment should exceed 5 $\mu_B$, for which
half metallicity is lost.

In view of recent experimental success to grow such compounds, and of
their relevance to the field of spintronics, we believe that our work
will not only add to the understanding of these systems, but will also
contribute to the future realization of spintronics devices.

\begin{acknowledgments}
  Financial support from the RT Network of {\em Computational
    Magnetoelectronics} (contract RTN1-1999-00145) of the European
  Commission is greatfully acknowledged.  The authors would like to
  thank Professor P.~H.~Dederichs for helpful discussions and for a
  critical reading of the manuscript. Motivating discussions with
  Professor H.~Katayama-Yoshida and Dr.~K.~Sato are also gratefully
  acknowledged.
\end{acknowledgments}


\begin{thebibliography}{99}
\bibitem{Janak77}
O.~Gunnarsson, J.~Phys.~F {\bf 6}, 587 (1976); J.~F.~Janak,
Phys.~Rev.~B {\bf 16}, 255 (1977). In the latter reference the
exchange integral is by definition half the one used here.
\end{thebibliography}
\end{document}